\documentclass[a4paper,11pt]{article}
\usepackage[dvips]{graphicx}
\usepackage{amssymb}
\usepackage{amsmath}

\usepackage{cite}

\begin{document}

\title{De Sitter Cosmic Strings and Supersymmetry}
\author{V. K. Oikonomou\thanks{
Vasilis.Oikonomou@mis.mpg.de}\\
Max Planck Institute for Mathematics in the Sciences\\
Inselstrasse 22, 04103 Leipzig, Germany} \maketitle

\begin{abstract}
We study massive spinor fields in the geometry of a straight
cosmic string in a de Sitter background. We find a hidden $N=2$
supersymmetry in the fermionic solutions of the equations of
motion. We connect the zero mode solutions to the heat-kernel
regularized Witten index of the supersymmetric algebra.
\end{abstract}

\section*{Introduction}

Cosmic strings are topological defects that could have been
created during cosmological phase transitions in the early
universe \cite{vilenkin1}. Beyond all question, these defects
cannot be responsible for the primordial density perturbations in
the Cosmic Microwave Background radiation. However the interest in
cosmic strings has been renewed since they can be related to a
number of physical phenomena \cite{vilenkin2,brane,saharian}.

\noindent Cosmic strings cause gravitational phenomena due to the
extremely large mass per unit length these have. Therefore, even
if the string is a straight line, it affects drastically spacetime
around it. For a straight string, spacetime is flat except from a
small deficit angle where curvature has a conical singularity.
This angle can cause many astrophysical effects, such as doubling
images of distant objects (for example quasars), or even cause
gravitational lensing. Furthermore, wiggly cosmic strings could
explain structure formation. In addition, emission of
gravitational waves can be explained with the help of cosmic
strings.

\noindent Astrophysical phenomena could find their origin in
cosmic string theories. Examples of such astrophysical phenomena
are high energy cosmic rays in our galaxy and primordial galactic
magnetic fields. The later are related to superconducting cosmic
strings \cite{witten} from which ultra-high currents (that consist
of charged localized matter) are emitted.

\noindent In our study we shall use straight cosmic strings.
Although real strings are not straight, they can be thought to be
chains of small straight segments. Thus, calculations for straight
strings can be very useful. The background we shall use is that of
de Sitter spacetime. This spacetime is a maximal symmetric
solution to Einstein's equations, with $R^1\times S^3$ topology.
Due to the high symmetry that this gravitational background has,
many problems are exactly solvable. Therefore studies of such
spacetimes can shed light to more difficult problems.

\noindent We focus our study on massive fermions around cosmic
strings in de Sitter spacetime. We find a hidden $N=2$
supersymmetric quantum mechanics algebra underlying the system and
we relate the number of zero mode solutions of the equation of
motion to the Witten index of the supersymmetric algebra.

\noindent The connection of a supersymmetric algebra with a Dirac
fermionic system is not accidental. It is known that Dirac
operators $H$ can be split into even and odd parts, that is,
$H=H_{+}+H_{-}$ (with $H_+$ denoting the even part and $H_-$
denoting the odd part), and this fact is actually closely related
with the general notion of supersymmetry \cite{thaller}.
Particularly, when $H_{+}$ anti-commutes with $H_{-}$, then $H$ is
called a supersymmetric Dirac operator. Supersymmetric quantum
mechanics in Dirac and gauge theories related has been studied in
recent works. For the connection of extra dimensional gauge models
and $N=2$ supersymmetric quantum mechanics algebra ($N=2$ susy QM
thereafter) see \cite{japs}. Also localized fermions around
superconducting cosmic strings are connected with $N=2$ susy QM
algebra, see \cite{oikonomoustrings}.

\medskip

This article is organized as follows: First we briefly review the
$N=2$ supersymmetric quantum mechanics algebra, next we derive the
equations of motion of the fermions around the cosmic strings and
relate the fermionic system with the $N=2$ susy QM algebra.
Finally we present the conclusions along with a discussion.

\section*{Supersymmetric Quantum Mechanics}

We briefly review the $N=2$ supersymmetric quantum mechanics
\cite{susyqm,susyqm1} algebra, relevant to our analysis.

\noindent Consider a quantum system, described by the self-adjoint
Hamiltonian operator $H$ and characterized by the set of
self-adjoint operators $\{Q_1,...,Q_N\}$. The quantum system is
supersymmetric, if,
\begin{equation}\label{susy1}
\{Q_i,Q_j\}=H\delta_{i{\,}j}
\end{equation}
with $i=1,2,...N$. The $Q_i$  are the supercharges and the
Hamiltonian ``$H$" is called supersymmetric (from now on ``susy")
Hamiltonian. The algebra (\ref{susy1}) constitutes the N-extended
supersymmetry. Owing to the anti-commutativity one has,
\begin{equation}\label{susy3}
H=2Q_1^2=2Q_2^2=\ldots =2Q_N^2=\frac{2}{N}\sum_{i=1}^{N}Q_i^2.
\end{equation}
A supersymmetric quantum system is said to have unbroken
supersymmetry, if its ground state vanishes, that is $E_0=0$. When
$E_0>0$, susy is said to be broken.

\noindent In order susy is unbroken, the ground states that belong
in the Hilbert space of all the eigenstates, must be annihilated
by the supercharges,
\begin{equation}\label{s1}
Q_i |\psi_0\rangle=0.
\end{equation}

\subsection*{$N=2$ supersymmetric quantum mechanics algebra}

The $N=2$ algebra consists of two supercharges $Q_1$ and $Q_2$ and
a Hamiltonian $H$, which obey,
\begin{equation}\label{sxer2}
\{Q_1,Q_2\}=0,{\,}{\,}{\,}H=2Q_1^2=2Q_2^2=Q_1^2+Q_2^2
\end{equation}
We introduce the operator,
\begin{equation}\label{s2}
Q=\frac{1}{\sqrt{2}}(Q_{1}+iQ_{2})
\end{equation}
and the adjoint,
\begin{equation}\label{s255}
Q^{\dag}=\frac{1}{\sqrt{2}}(Q_{1}-iQ_{2})
\end{equation}
The above two operators satisfy,
\begin{equation}\label{s23}
Q^{2}={Q^{\dag}}^2=0
\end{equation}
and are related to the Hamiltonian as,
\begin{equation}\label{s4}
\{Q,Q^{\dag}\}=H
\end{equation}
The Witten parity, $W$, for a $N=2$ algebra is defined as,
\begin{equation}\label{s45}
[W,H]=0
\end{equation}
and
\begin{equation}\label{s5}
\{W,Q\}=\{W,Q^{\dag}\}=0
\end{equation}
Also $W$ satisfies,
\begin{equation}\label{s6}
W^{2}=1
\end{equation}
By using $W$, the Hilbert space $\mathcal{H}$ of the quantum
system is spanned to positive and negative Witten parity subspaces
which are defined as,

\begin{equation}\label{shoes}
\mathcal{H}^{\pm}=P^{\pm}\mathcal{H}=\{|\psi\rangle :
W|\psi\rangle=\pm |\psi\rangle \}
\end{equation}
Therefore, the Hilbert space $\mathcal{H}$ is decomposed into the
eigenspaces of $W$, hence $\mathcal{H}=\mathcal{H}^+\oplus
\mathcal{H}^-$. Each operator acting on the vectors of
$\mathcal{H}$ can be represented by $2N\times 2N$ matrices. We use
the representation:
\begin{equation}\label{s7345}
W=\bigg{(}\begin{array}{ccc}
  I & 0 \\
  0 & -I  \\
\end{array}\bigg{)}
\end{equation}
with $I$ the $N\times N$ identity matrix. Bring to mind that
$Q^2=0$ and $\{Q,W\}=0$, hence the supercharges are of the form,
\begin{equation}\label{s7}
Q=\bigg{(}\begin{array}{ccc}
  0 & A \\
  0 & 0  \\
\end{array}\bigg{)}
\end{equation}
and
\begin{equation}\label{s8}
Q^{\dag}=\bigg{(}\begin{array}{ccc}
  0 & 0 \\
  A^{\dag} & 0  \\
\end{array}\bigg{)}
\end{equation}
which imply,
\begin{equation}\label{s89}
Q_1=\frac{1}{\sqrt{2}}\bigg{(}\begin{array}{ccc}
  0 & A \\
  A^{\dag} & 0  \\
\end{array}\bigg{)}
\end{equation}
and also,
\begin{equation}\label{s10}
Q_2=\frac{i}{\sqrt{2}}\bigg{(}\begin{array}{ccc}
  0 & -A \\
  A^{\dag} & 0  \\
\end{array}\bigg{)}
\end{equation}
The $N\times N$ matrices $A$ and $A^{\dag}$, are generalized
annihilation and creation operators. The action of $A$ is defined
as $A: \mathcal{H}^-\rightarrow \mathcal{H}^+$ and that of
$A^{\dag}$ as, $A^{\dag}: \mathcal{H}^+\rightarrow \mathcal{H}^-$.
In the representation (\ref{s7345}), (\ref{s7}), (\ref{s8}) the
Hamiltonian $H$, can be cast in a diagonal form \footnote{The
diagonal form of a Hamiltonian is most welcome, since the spectral
analysis of the Hamiltonian can be reduced to the analysis of
simpler operators. However if an off diagonal form is preferred,
one can perform a Foldy-Wouthuysen transformation
\cite{thaller}.},
\begin{equation}\label{s11}
H=\bigg{(}\begin{array}{ccc}
  AA^{\dag} & 0 \\
  0 & A^{\dag}A  \\
\end{array}\bigg{)}
\end{equation}
Therefore the total supersymmetric Hamiltonian $H$, consists of
two superpartner Hamiltonians,
\begin{equation}\label{h1}
H_{+}=A{\,}A^{\dag},{\,}{\,}{\,}{\,}{\,}{\,}{\,}H_{-}=A^{\dag}{\,}A
\end{equation}
We define the operator $P^{\pm}$. The eigenstates of $P^{\pm}$,
denoted as $|\psi^{\pm}\rangle$  are called positive and negative
parity eigenstates which satisfy,
\begin{equation}\label{fd1}
P^{\pm}|\psi^{\pm}\rangle =\pm |\psi^{\pm}\rangle
\end{equation}
Using the representation (\ref{s7345}), the parity eigenstates are
represented in the form,
\begin{equation}\label{phi5}
|\psi^{+}\rangle =\left(%
\begin{array}{c}
  |\phi^{+}\rangle \\
  0 \\
\end{array}%
\right)
\end{equation}
and also,
\begin{equation}\label{phi6}
|\psi^{-}\rangle =\left(%
\begin{array}{c}
  0 \\
  |\phi^{-}\rangle \\
\end{array}%
\right)
\end{equation}
with $|\phi^{\pm}\rangle$ $\epsilon$ $H^{\pm}$.

\noindent  In order to have unbroken supersymmetry, there must be
at least one state in the Hilbert space (we denote it as
$|\psi_{0}\rangle$ ) with vanishing energy eigenvalue, that is
$H|\psi_{0}\rangle =0$. This implies that $Q|\psi_{0}\rangle =0$
and $Q^{\dag}|\psi_{0}\rangle =0$. For a ground state with
negative parity,
\begin{equation}\label{phi5}
|\psi^{-}_0\rangle =\left(%
\begin{array}{c}
  0 \\
  |\phi^{-}_{0}\rangle \\
\end{array}%
\right)
\end{equation}
this would imply that $A|\phi^{-}_{0}\rangle =0$, while for a
positive parity ground state,
\begin{equation}\label{phi6s6}
|\psi^{+}_{0}\rangle =\left(%
\begin{array}{c}
  |\phi^{+}_0\rangle \\
  0 \\
\end{array}%
\right)
\end{equation}
it would imply that $A^{\dag}|\phi^{+}_{0}\rangle =0$. A ground
state can either have positive or negative Witten parity.
Nevertheless, when the ground state is degenerate, both cases can
occur. When $E\neq 0$, the number of positive parity eigenstates
is equal to the negative parity eigenstates. Yet, this does not
hold for the zero modes. Zero modes are fully described by the
Witten index. Let $n_{\pm}$ be the number of zero modes of
$H_{\pm}$ in the subspace $\mathcal{H}^{\pm}$. For a finite number
of zero modes (which implies the operator $A$ is Fredholm
\footnote{If an operator $A$ is Fredholm, this implies that it has
discrete spectrum. In addition the Fredholm property is ensured if
$\mathrm{dim{\,}ker}A<\infty$. Equivalently if an operator is
trace-class, then it is by definition Fredholm. For a extensive
analysis on these issues, see \cite{thaller}}), $n_{+}$ and
$n_{-}$, the quantity,
\begin{equation}\label{phil}
\Delta =n_{-}-n_{+}
\end{equation}
is called the Witten index. When the Witten index is non-zero
integer, supersymmetry is unbroken and if it is zero, it is not
clear whether supersymmetry is broken. If $n_{+}=n_{-}=0$
supersymmetry is obviously broken, but if $n_{+}= n_{-}\neq 0$
supersymmetry is not broken.

\noindent The Fredholm index of the operator $A$ is closely
related to the Witten index. The former is defined as,
\begin{equation}\label{ker}
\mathrm{ind} A = \mathrm{dim}{\,}\mathrm{ker}
A-\mathrm{dim}{\,}\mathrm{ker} A^{\dag}=
\mathrm{dim}{\,}\mathrm{ker}A^{\dag}A-\mathrm{dim}{\,}\mathrm{ker}AA^{\dag}
\end{equation}
Indeed we have,
\begin{equation}\label{ker1}
\Delta=\mathrm{ind} A=\mathrm{dim}{\,}\mathrm{ker}
H_{-}-\mathrm{dim}{\,}\mathrm{ker} H_{+}
\end{equation}
If the operator $A$ is not Fredholm, then the Witten index is not
defined as in (\ref{phil}) and (\ref{ker1}). However there exists
a heat-kernel regularized index, both for the operator $A$ (which
we denote $\mathrm{ind}_tA$) and for the Witten index, $\Delta_t$.
The regularized index for the operator $A$ is defined as:
\begin{equation}\label{heatkerw}
\mathrm{ind}_tA=\mathrm{tr}e^{-tA^{\dag}A}-\mathrm{tr}e^{-tAA^{\dag}}
\end{equation}
with $t>0$ and the trace is taken over the eigenfunctions
corresponding to the zero modes of $A$. The regularized Witten
index is equal to:
\begin{equation}\label{hetreg}
\Delta_t=\lim_{t\rightarrow \infty}\mathrm{ind}_tA
\end{equation}
In the following we shall extensively use the definitions we gave
and the notation we used in this section.

\section*{Fermions Around de Sitter Cosmic Strings and $N=2$ SUSY QM}

Consider and infinitely long straight cosmic string. Due to the
cylindrical symmetry, the line element in cylindrical coordinates
$(r,\phi,z)$ is (we use the notation of \cite{saharian}):
\begin{equation}\label{metric}
\mathrm{d}s^2=g_{ \mu
\nu}\mathrm{d}x^{\mu}\mathrm{d}x^{\nu}=\mathrm{d}t^2-e^{2t/a}(\mathrm{d}r^2+r^2\mathrm{d}\phi^2+\mathrm{d}z^2),
\end{equation}
with $r\geq 0$, $-\infty <z <\infty$. The points $\phi$ and
$\phi_{0}$ on the hypersurface $z=const$ and $r=const$ are
considered to be identical. The parameter $a$ is equal to
$a=\sqrt{3/\Lambda}$, with $\Lambda$ the cosmological constant.





\noindent The Dirac equation in the curved spacetime reads
\cite{Jost,Nakahara,eguchi,saharian},
\begin{equation}\label{spinor}
i\gamma^{\mu}\nabla_{\mu}\psi-m\psi=0,{\,}{\,}{\,}{\,}\nabla_{\mu}=\partial_{\mu}+\Gamma_{\mu}
\end{equation}
where $\gamma^{\mu}$ and $\Gamma^{\mu}$, stand for the curved
spacetime gamma matrices and spin connection respectively. Using
the vierbeins $e^{\mu}_{(a)}$, we can connect the curved spacetime
gamma matrices, to the flat spacetime ones:
\begin{equation}\label{gammadir}
\gamma^{\mu}=e^{\mu }_{(a)}\gamma^{(a)}, {\,}{\,}\Gamma_{\mu
}=\frac{1}{4}\gamma^{(a)}\gamma^{(b)}e^{\nu
}_{(a)}e_{{(b)}{\,}{\nu ;\mu }}
\end{equation}
with $g^{\mu \nu}=e^{\mu}_{(a)}e^{\nu}_{(b)}\eta^{ab}$ and
$\mu=0,1,2,3,4$. The flat space spacetime Dirac matrices are,
\begin{equation}\label{gammaflat}
\gamma^{(0)}=\bigg{(}\begin{array}{ccc}
  1 & 0 \\
  0 & -1  \\
\end{array}\bigg{)},{\,}{\,}{\,}{\,}{\,}\gamma^{(a)}=\bigg{(}\begin{array}{ccc}
  0 & \sigma_{\alpha} \\
  -\sigma_{\alpha} & 0  \\
\end{array}\bigg{)}{\,}{\,}{\,}\alpha=1,2,3
\end{equation}
with $\sigma_i$ the Pauli matrices. Using the vierbeins,
\begin{equation}\label{connections}
e^{\mu}_{(a)}=e^{-t/a} \left ( \begin{array}{cccc}
  e^{t/a} & 0 & 0 & 0 \\
  0 & \cos(q \phi) & -\sin(q\phi)/r & 0 \\
0 & \sin(q\phi) & \cos(q\phi) & 0 \\
0 & 0 & 0 & 1  \\
\end{array} \right)
\end{equation}
with $q=2\pi/\phi_0$, the curved spacetime gamma matrices can be
written as, $\gamma^{0}=\gamma^{(0)}$ and
\begin{equation}\label{gammalast}
\gamma^{i}=e^{t/a}\left (\begin{array}{ccc}
  0 & \beta^i \\
  -\beta^i & 0  \\
\end{array}\right )
\end{equation}
The matrices $\beta^i$ are equal to,
\begin{equation}\label{beta1}
\beta^{1}=\left (\begin{array}{ccc}
  0 & e^{-iq\phi} \\
  e^{iq\phi} & 0  \\
\end{array}\right ){\,}{\,}{\,}{\,}{\,} \beta^{2}=-\frac{i}{r}\left (\begin{array}{ccc}
  0 & e^{-iq\phi} \\
  -e^{iq\phi} & 0  \\
\end{array}\right ){\,}{\,}{\,}{\,}{\,}
\beta^{3}=\bigg{(}
\begin{array}{ccc}
  1 & 0 \\
  0 & -1 \\
\end{array} \bigg{)}
\end{equation}
Finally the spin connections are,
\begin{equation}\label{Gammabig}
\Gamma_0=0,{\,}{\,}\Gamma_i=-\frac{1}{2a}\gamma^0\gamma_i+\frac{1-q}{2}\gamma^{(1)}\gamma^{(2)}\delta_i^2,
{\,}{\,}{\,}i=1,2,3
\end{equation}
Decomposing the spinor $\psi$ in the following form,
\begin{equation}\label{bispinor}
\psi=\left(%
\begin{array}{c}
  \phi \\
  \chi \\
\end{array}%
\right)
\end{equation}
the fermionic equations of motion around a cosmic string, in de
Sitter background, can be cast as:
\begin{align}\label{eqmotion}
&D_{+}\phi+\left(\beta^l\partial_l+\frac{1-q}{2r}\beta^1\right
)\chi=0\\ \notag
&D_{-}\chi+\left(\beta^l\partial_l+\frac{1-q}{2r}\beta^1\right
)\phi=0
\end{align}
In the above equations, $D_{\pm}$, stand for:
\begin{equation}\label{operators}
D_{\pm}=\partial_r-\frac{1}{\tau}\left(\frac{3}{2}\pm ima \right )
\end{equation}
with $\tau=-ae^{t/a}$, $-\infty<\tau <0$. Note that
$D_{+}=D_{-}^*$, which means the complex conjugate of $D_{+}$ is
$D_{-}$. Also the operator
$\mathcal{B}=\beta^l\partial_l+\frac{1-q}{2r}\beta^1$ is
self-adjoint. Using the operators $D_{\pm}$ and also
$\mathcal{B}$, we can form an $N=2$ susy QM algebra. Indeed, we
define,
\begin{equation}\label{dmatrix1}
D=\left(%
\begin{array}{cc}
  D_{+} & \beta^i\partial_i+\frac{1-q}{2r}\beta^1
 \\
 \beta^i\partial_i+\frac{1-q}{2r}\beta^1
 & D_{-} \\
\end{array}%
\right)
\end{equation}
acting on,
\begin{equation}\label{wee33}
|\phi^{-}\rangle=\left(%
\begin{array}{c}
  \phi \\
  \chi \\
\end{array}%
\right).
\end{equation}
Upon taking the adjoint we obtain,
\begin{equation}\label{dmatrix23}
D^{\dag}=\left(%
\begin{array}{cc}
  D_{-} & \beta^i\partial_i+\frac{1-q}{2r}\beta^1
 \\
 \beta^i\partial_i+\frac{1-q}{2r}\beta^1
 & D_{+} \\
\end{array}%
\right)
\end{equation}
acting on,
\begin{equation}\label{cartman}
|\phi^{+}\rangle=\left(%
\begin{array}{c}
  \chi \\
  \phi \\
\end{array}%
\right)
\end{equation}
The equation $D|\phi^{-}\rangle =0$ (we used the notation of
(\ref{phi5}) for reasons that will become clear shortly) yields
the solutions of the equation of motion (\ref{eqmotion}). Hence,
it is easy to see that the zero modes of the operator $D$
correspond to the solutions of the equation of motion
(\ref{eqmotion}). Notice that, the zero modes of the operator
$D^{\dag}$ are $\chi$ and $\phi$, as can be easily checked
(actually the equation $D^{\dag}|\phi^{+}\rangle =0$ yields the
equations of motion (\ref{eqmotion})). Using the operators $D$ and
$D^{\dag}$, we can define the supercharges $Q$ and $Q^{\dag}$,
\begin{equation}\label{wit2}
Q=\bigg{(}\begin{array}{ccc}
  0 & D \\
  0 & 0  \\
\end{array}\bigg{)}, {\,}{\,}{\,}{\,}{\,}Q^{\dag}=\bigg{(}\begin{array}{ccc}
  0 & 0 \\
  D^{\dag} & 0  \\
\end{array}\bigg{)}
\end{equation}
Also the Hamiltonian of the system can be written in terms of $D$
and $D^{\dag}$, in the following diagonal form,
\begin{equation}\label{wit4}
H=\bigg{(}\begin{array}{ccc}
  DD^{\dag} & 0 \\
  0 & D^{\dag}D  \\
\end{array}\bigg{)}
\end{equation}
It is obvious that the above matrices obey, $\{Q,Q^{\dag}\}=H$,
$Q^2=0$, ${Q^{\dag}}^2=0$, $\{Q,W\}=0$, $W^2=I$ and $[W,H]=0$.
Thus we can see that an $N=2$ susy QM algebra underlies the
fermionic system. This property is it self particularly useful,
especially for spectral problems of Dirac fields around defects
(straight on deformed). Let us now see what are the implications
of supersymmetry in the fermionic system and it's zero mode
solutions. We are particularly interested in the zero mode
solutions but, we shall also discuss at the end of this section
the implications of supersymmetry on the eigenfunctions of the
Hamiltonian with $E\neq 0$.

\noindent The operator $D$ is not Fredholm because it has not
discrete spectrum. This can be easily seen from equation
(\ref{eqmotion}). Indeed it can be written as:
\begin{equation}\label{bigequation}
\left
(\partial_r^2+\frac{1}{r}\partial_r+\frac{1}{r^2}\partial_{\phi}^2+\partial_r^2+\frac{q-1}{r}\beta^1\beta^2\partial_{\phi}-\frac{(q-1)^2}{4r^2}-D_{+}D_{-}\right
)\phi =0
\end{equation}
The solution reads,
\begin{equation}\label{onesolution}
\psi_{\sigma}(x)\sim \left ( \begin{array}{c}
 C_1 H_{1/2-ima}^{(1)}(\gamma \eta )J_{\beta_1}(\lambda r) \\
  C_2 H_{1/2-ima}^{(1)}(\gamma \eta )J_{\beta_1}(\lambda r)e^{iq\phi} \\
  C_3 H_{-1/2-ima}^{(1)}(\gamma \eta )J_{\beta_1}(\lambda r)\\
  C_4 H_{-1/2-ima}^{(1)}(\gamma \eta )J_{\beta_1}(\lambda r)e^{iq\phi}\\
\end{array}   \right )
\end{equation}
with ``$\sigma$" characterizing the quantum numbers of the system,
two of which continuously vary from zero to infinity ($\lambda$
and $k$). For a complete analysis on the solutions see
\cite{saharian}. Accordingly, the definitions
(\ref{heatkerw}) and (\ref{hetreg}) for the regularized indices
hold.

\noindent Since the two equations $D|\phi^{-}\rangle =0$ and
$D^{\dag}|\phi^{+}\rangle=0$ have the same solutions, namely
$\chi$ and $\phi$, this implies that
$\mathrm{ker}D=\mathrm{ker}D^{\dag}$, which in turn implies
$\mathrm{ker}DD^{\dag}=\mathrm{ker}DD^{\dag}$. Hence the operators
$e^{-tD^{\dag}D}$ and $e^{-tDD^{\dag}}$ have the same trace, that
is $\mathrm{tr}e^{-tD^{\dag}D}=\mathrm{tr}e^{-tDD^{\dag}}$. This
means that the regularized index of the operator $D$ is zero and
hence the regularized Witten index is zero. Therefore,
supersymmetry is unbroken. In order to further clarify this point,
recall that supersymmetry is unbroken in two cases: when the
Witten index is non-zero integer and if it is zero, but the number
of zero modes satisfy $n_{+}= n_{-}\neq 0$. If on the contrary
$n_{+}=n_{-}=0$ supersymmetry is obviously broken. In the
non-Fredholm operator case, the numbers $n_{+}$ and $n_{-}$, that
is, the zero modes of the operator $D$ and $D^{\dag}$
respectively, are replaced by $\mathrm{ker}D$ and
$\mathrm{ker}D^{\dag}$. Since
$\mathrm{ker}D=\mathrm{ker}D^{\dag}\neq 0$, we conclude that just
as in the Fredholm operators case, supersymmetry is unbroken.
Clearly this corresponds to the case we described below equation
(\ref{onesolution}).

\noindent The relation $D|\phi^{-}\rangle=0$ (we use the notation
of relations (\ref{phi6}) and (\ref{phi5})) implies
$Q|\psi^{-}\rangle=0$. This means that the ground state $\Psi^{-}$
is actually a negative parity eigenstate. In such a way, the
negative Witten parity zero mode eigenstate $|\psi^{-}_0\rangle$,
of the Hamiltonian $H^-$, is:
\begin{equation}\label{neg}
|\psi^{-}_0\rangle=\left(%
\begin{array}{c}
  0\\
  0\\
  \phi \\
  \chi \\
\end{array}%
\right)
\end{equation}
In the same vain, the positive parity zero mode eigenstate is,
\begin{equation}\label{negkar}
|\psi^{+}_0\rangle =\left(
\begin{array}{c}
  \chi \\
  \phi \\
   0 \\
   0 \\
\end{array}
\right)
\end{equation}
Therefore, owing to supersymmetry, the $\chi$ and $\phi$
components of the Dirac spinor constitute the positive and
negative parity solutions of the $N=2$ supersymmetric system.

\noindent Let us see the implications
of supersymmetry on the eigenfunctions of the Hamiltonian with
$E\neq 0$. The Hamiltonians $H_+$ and $H_-$, are known to be
isospectral for eigenvalues different from zero
\cite{thaller,susyqm}, that is,
\begin{equation}\label{isosp}
\mathrm{spec}(H_{+})\setminus \{0\}=\mathrm{spec}(H_{-})\setminus
\{0\}
\end{equation}
In addition, the following relations hold,
\begin{equation}\label{positivee}
Q|\psi^{-}_0\rangle=\sqrt{E}|\psi^{+}_0\rangle{\,}{\,}{\,}\mathrm{and}{\,}{\,}{\,}Q^{\dag}|\psi^{+}_0\rangle=\sqrt{E}|\psi^{-}_0\rangle,
\end{equation}
with $E$ the common eigenvalues of the Hamiltonians $H_+$ and
$H_-$. In turn these imply,
\begin{equation}\label{pose}
D|\phi^{-}\rangle=\sqrt{E}|\phi^{+}\rangle
{\,}{\,}{\,}\mathrm{and}{\,}{\,}{\,}D^{\dag}|\phi^{+}\rangle=\sqrt{E}|\phi^{-}\rangle.
\end{equation}

Apart from these, there are many interesting mathematical
properties that the supersymmetric system possesses, but these are
beyond the scope of this article \footnote{For example the Krein's
spectral function $\xi(\lambda)$, which is equal to zero because
the Witten index is zero.}.

We have to note that supersymmetry of any kind around topological defects and in
various gravitational backgrounds, has been studied in many works,
for example \cite{schwarzchild,monopole,taub}. In reference
\cite{schwarzchild}, there was found that there is a close
connection between supersymmetric quantum mechanics and the
mechanics of a spinning Dirac particle moving in Schwarzschild
spacetime. In addition in reference \cite{monopole} it was found
that a fermion in a monopole field possesses a rich supersymmetry
structure. The connection of supersymmetry with the Taub-NUT
spacetime was studied in \cite{taub}.

\noindent Studies involving de Sitter space are of particular
importance. Indeed there has been a great deal of work on de
Sitter space solutions in supergravity and string theory (see
\cite{linde} and references therein). The connection of de Sitter
space and supercritical string theory was presented in
\cite{critstring}. In addition, de Sitter space is known to have
finite entropy. De Sitter entropy can be understood as the number
of  degrees of freedom in a quantum mechanical dual \cite{horava}.
Moreover, de Sitter space is closely connected with $N=2$
supersymmetry \cite{renata} and also these two are connected with
hybrid inflation solutions. Furthermore, $N=4$ supersymmetric
quantum mechanics is very closely related to the description of
particle dynamics in de Sitter space.

\noindent Obviously spacetime supersymmetry and supersymmetric
quantum mechanics are not the same, nevertheless the connection is
profound, since extended (with $N=4,6...$) supersymmetric quantum
mechanics models describe the dimensional reduction to one
(temporal) dimension of $N=2$ and $N=1$ Super-Yang Mills models
\cite{pashev}.

\noindent The interconnection of supersymmetric theories is an
interesting mathematical property that is closely connected with
the construction of superconformal quantum mechanical models
\cite{pashev}. In addition, a number of physical applications,
renders studies of supersymmetric quantum mechanics models to be
of valuable importance (due to the simplicity these have), such as
low energy dynamics of black hole models, see \cite{pashev}.

\noindent Localized fermion solutions around vortices and other
defects frequently appear in superconductor studies. Thus in view
of a micro-description of superconductors, through holographic
superconductors \cite{hol}, the existence of a supersymmetric
quantum mechanic algebra could be of particular importance.

\noindent From the above, it is clear that supersymmetric quantum
mechanics is a very useful tool to make complex field theories
more easy to handle. In addition, studying fermionic solutions
around defects in de Sitter space is very interesting, since these
problems may be reductions of more evolved problems.

\noindent But there is another valuable property of the
supersymmetric quantum mechanics supercharges that we did not
mentioned. The set of the $N=2$ supersymmetric quantum mechanics
supercharges are invariant under an R-symmetry. Furthermore, the
Hamiltonian is invariant under this symmetry \cite{susyqm}.
Particularly, the real superalgebra (\ref{susy1}) and
(\ref{sxer2}) is invariant under the transformation,
\begin{equation}\label{rsymetry}
\left (\begin{array}{c}
  Q'_1 \\
   Q'_2  \\
\end{array}\right )=\left (\begin{array}{ccc}
  \cos a & \sin a \\
  -\sin a & \cos a  \\
\end{array}\right ) \cdot \left ( \begin{array}{c}
  Q_1 \\
  Q_2 \\
\end{array} \right )
\end{equation}
Correspondingly the complex supercharges $Q$ and $Q^{\dag}$ are
transformed under a global $U(1)$ transformation:
\begin{equation}\label{supercharges}
Q^{'}=e^{-ia}Q, {\,}{\,}{\,}{\,}{\,}{\,}{\,}{\,}
{\,}{\,}Q^{'\dag}=e^{ia}Q^{\dag}
\end{equation}
This R-symmetry is a symmetry of the Hilbert states corresponding
to the spaces $\mathcal{H}^{+}$ and $\mathcal{H}^{-}$. Furthermore
the two spaces can have different transformation parameters. To
make this clear, let $\psi^{+}$ and $\psi^{-}$ denote the Hilbert
states corresponding to the spaces $\mathcal{H}^{+}$ and
$\mathcal{H}^{-}$. Then the $U(1)$ transformation of the states
is,
\begin{equation}
\psi^{'+}=e^{-i\beta_{+}}\psi^{+},
{\,}{\,}{\,}{\,}{\,}{\,}{\,}{\,}
{\,}{\,}\psi^{'-}=e^{-i\beta_{-}}\psi^{-}
\end{equation}
It is clear that the parameters $\beta_{+}$ and $\beta_{-}$ are
different global parameters. Consistency with relation
(\ref{supercharges}) requires that $a=\beta_{+}-\beta_{-}$. We
must note that a global $U(1)$ symmetry is expected around a
cosmic string \cite{meierovich}. Furthermore the breaking of a
local $U(1)$ leaves a global $U(1)$ in cosmic string models
\cite{vilenkin1}. Thus this global $U(1)$ symmetry is really
interesting. Nevertheless we must further study whether there is a
connection between the susy R-symmetry and the remnant global
$U(1)$ symmetry of phenomenological models around strings. It
would then be interesting to study more realistic models, but this
is beyond the scopes of this article.

\noindent The cylindrical line element (\ref{metric}) has played a
crucial role in our analysis. An important question that is raised
is whether the outcomes of the de-Sitter cosmic string framework
hold for more general spacetimes. The answer in this question is
not so trivial and we should be very cautious in generalizing our
results. We have applied our results to cosmic strings in flat
spacetime and also in anti-de Sitter spacetimes and no
supersymmetry algebra underlies the fermionic system in these two
cases. Therefore such an analysis must be carried away with great
detail and caution and certainly would be invaluable. Also
possible generalizations to include bosonic or vector
configurations could be interesting. But this analysis is beyond
the purposes of this article.

\noindent Before closing this section we must mention some
supersymmetric quantum mechanics developments, relevant to our
work. The
supersymmetry studied in this article is similar to what takes
place in some periodic systems where $n_+=n_-=1\neq 1$, that is,
susy is exact though the Witten index is equal to zero
\cite{pluskai1,pluskai2,pluskai3}.

\noindent The peculiarity of the periodic systems
\cite{pluskai1,pluskai2,pluskai3} is that besides an $N=2$
supersymmetry, these systems possess much more rich structure
(related to their special nature being finite gap systems) which
is related to the existence of a hidden supersymmetry. Since in
our case, our system is characterized by the same property
$n_+=n_-\neq 0$, maybe it also possess a hidden, bosonized
supersymmetry. Hidden, bosonized supersymmetry  is related to the
existence of a grading operator, which has a nonlocal nature
(reflection operator) \cite{pluskai4,pluskai5,pluskai6}. We hope
we pursuit further these issues in a future publication.

\section*{Conclusions}

In this paper we found a hidden $N=2$ susy QM algebra, underlying
the fermionic solutions of the Dirac equation around a straight
cosmic string in de Sitter spacetime. The negative and positive
Witten parity states of the susy algebra can be written in terms
of the fermionic solutions of the equations of motion. The Witten
index is zero and additionally the number of zero modes for the
two supercharges are equal. We therefore concluded that the system
has unbroken supersymmetry. It was tempting to check whether this
susy algebra underlies the fermion system around a cosmic string
but with a flat background. It turns out that there is no
supersymmetry in the flat case. It stands to reason to argue that,
the presence of spacetime curvature is responsible for the susy
structure in the de Sitter fermion system. However this must not
be the case, since in the superconducting string case, the
background spacetime is also flat, still, an $N=2$ susy QM algebra
is present \cite{oikonomoustrings}. We hope to comment on this
issues in the future.

\noindent Before closing we must note that there is a mathematical
property stemming from $N=2$ susy QM, that can be very useful
perhaps when studying perturbations of the metric around the
string, or in the case the string is not straight (also perhaps in
the case of an electromagnetic field around the string. However we
must further study the last case to be sure. We hope to do so in a
future work). If these effects can be described by a matrix $C$,
which anti-commutes with the Witten operator and is  also
symmetric \footnote{Also the operator $Ce^{-D^{\dag}D}$ must be
trace-class}, then,
\begin{equation}\label{indperturbation}
\mathrm{ind}_{t}(D+C)=\mathrm{ind}_{t}D
\end{equation}
This means that there is a correspondence between the solutions of
the equation \linebreak $(D+C)\psi=0$ and these of the equation
$D\psi=0$. This is very useful, regarding fermionic spectral
problems around one dimensional defects.

.

\end{document}